# Evaluation of Forward Fall on the Outstretched Hand Using MADYMO Human Body Model

Nader Rajaei, *Member, IEEE*, Saeed Abdolshah, *Member, IEEE*, Yasuhiro Akiyama, *Member, IEEE*, Yoji Yamada *Member, IEEE*, Shogo Okamoto *Member, IEEE*,

*Abstract*—Falls and their related injuries pose a significant risk to human health. One of the most common falls, the forward fall, frequently occurs among adults and the elderly. In this study, we propose using a human body model, developed using the MAthematical DYnamic MOdel (MADYMO) software, in place of human subjects, to investigate forward fall-related injuries. The MADYMO human body model is capable of simulating items that cannot be assessed on human subjects, such as human kinematics, human dynamics, and the possibility of injuries. In order to achieve our goal, a set of experiments was conducted to measure the impact force during a worst-case forward fall scenario (the outstretched hand position) for two short fall heights. Similar to the experimental design used on the human subjects, we generated a MADYMO human model. After performing the simulations, the results of the experiment on the human subjects and the MADYMO simulation model were compared. We demonstrated a significant correlation between the MADYMO simulation and the human subject experiments with respect to the magnitude and timing of the impact forces. Consequently, we validated the MADYMO human body model as a means to accurately assess forward fall-related injuries. Additionally, we compared the predicted results of a mathematical model with the MADYMO human body model. The MADYMO model is reliable and can demonstrate an accurate impact time. Therefore, we conclude that the MADYMO human model can be utilized as a reliable model to investigate forward fall-related injuries from a typical standing position.

## I. Introduction

In most cases, falls pose a serious risk to human health, especially for senior individuals. According to the World Health Organization report [1], falls are the second leading cause of unintentional injury and death. An initial study showed that approximately 11% of falls can cause a serious injury for older adults [2]. In addition, falls can impose a significant cost to the healthcare system. Recent investigations regarding fall-related injuries in the United States show that the direct cost of medical care for fatal and non-fatal fall injuries were approximately $616.5 million and $30.3 million in 2012, respectively. These costs increased to $637.5 million and $31.3 million in 2015 [3]. Hence, it is of great interest in the field of biomechanics to introduce new methods for preventing and/or reducing fall-related injuries [4]-[5].

One of the most common types of falls is the forward fall [6]. The forward fall mainly occurs due to an unexpected perturbation of balance or a failure in recovery strategies [7]

N. Rajaei, S. Abdolshah, Y. Akiyama, Y. Yamada and S. Okamoto are with Department of Mechanical System Engineering, Nagoya University, Nagoya, Japan (e-mail: nader.rajaei@mae.nagoya-u.ac.jp).

when tripping or slipping occurs during walking [6]-[8]. In addition, the forward fall commonly occurs in the workplace, where there is the potential interaction between humans and robots.

Individual often uses the upper extremities for protecting the head and trunk from impact when a forward fall occurs [6]-[9]. Hence, forward falls lead frequently to fractures of the upper extremities, such as the Colles' fracture (distal radius fracture) [10]-[12]. A recent investigation has indicated that approximately 40.5% of forward falls cause moderate or severe injuries to individuals above 65 years of age [13]. Thus, it is of critical importance to evaluate forward falls and their related injuries and to develop strategies for reducing these injuries.

Earlier studies have focused on measuring the magnitude of the impact force that causes a distal radius fracture [14]-[17]. The researchers performed their experiments by exerting a force on a cadaveric hand. For instance, an investigation by Frykman (1967) showed an average impact force of $2.26 \pm 0.01$ kN for 48 cadaveric forearms (25 women and 23 men of ages 65 and older) [14]. A similar protocol was utilized by Myers (1991 and 1993) and Spadaro (1994); they determined that wrist fractures occurred at impact forces of $3.39 \pm 0.88$ (Myers's first study), $1.78 \pm 0.65$ (Myers's second study) [14]-[15], and $1.64 \pm 0.98$ kN (Spadaro's study) [17].

Although the former studies indicated a range of impact forces that produce a radial bone fracture, these *in vitro* studies had limitations because the roles of the reactions from the upper extremity muscles were not considered.

Subsequent investigations on fall-related injuries were conducted using human subjects [11], [18]-[22]. In one of the first studies, Chiu and Robinovitch (1998) [18] designed a biomechanical experiment involving the worst-case forward fall (the outstretched hand position). They obtained the impact force when there was a short distance between the participant's hand and the ground. The results indicated that the impact forces consisted of an initial high magnitude peak ($f_{max1}$), occurring shortly after the hand contacts the ground, followed by a subsequent lower peak ($f_{max2}$). Then, they examined a strategy for reducing the impact force. They compared the magnitude of the impact force when the hand contacted the ground having an infinite material stiffness with that of the ground consisting of a compliant material (i.e. a compliant flooring). They found that decreasing the stiffness of the ground leads to a reduced first peak magnitude ($f_{max1}$) but did not show a large change in the magnitude of the second peak ($f_{max2}$) [19]. In addition, the magnitude of $f_{max1}$ was further reduced where a rigid layer is mounted on a compliant flooring [23].

Other studies also attempted to examine the human reaction during a forward fall [24] and several different arrest strategies for reducing the impact forces. One study investigated the effect of elbow flexion movements after contacting the ground [11]. The result showed that the action of elbow flexion decreased the first impact force and caused a delay in the timing of the peak. In another study, DeGoede et al., [21] investigated the magnitude of the impact force for three elbow angles, ranging from almost outstretched to almost fully bent. They found similar impact force peaks, including $f_{max1}$ and $f_{max2}$. In addition, they could specify that the magnitude of the impact forces were maximum when the elbow was extended to $\theta_{Elbow} = 174°$ (the outstretched hand position), whereas the forces decreased by approximately 40% at the reduced elbow angle.

Despite these efforts, the *in vivo* studies have also shown some limitations due to safety considerations for the subjects when falling from a standing position. To solve this issue, Chiu and Robinovitch (1998) [18] proposed biodynamic impact models. The proposed model was based on a simple two-degree-of-freedom spring-mass-damper system that was capable of explaining the nature of a forward fall at a large height and in the outstretched hand position.

This simple mathematical model predicted the applied forces to the hand during the forward fall. However, the model had some limitations due to the underprediction and overprediction of the time of the impact force [18]. Therefore, the model was not capable of demonstrating the precise hand displacement and energy absorption. In this study, we propose a human body model developed using the Mathematical Dynamic Models (MADYMO) software [25]-[26] to predict the biomechanical factors related to a forward fall. The MADYMO human model has been generally utilized for occupant safety studies in the automotive and transport industries. The software is mostly validated for various loading conditions to the model's head, shoulder, thorax, pelvis, and legs (details can be found in the MADYMO Human Models Manual [25]). The MADYMO human body model has an advanced numerical algorithm for simulating human kinematics and human dynamics and can evaluate the possibility of injuries that cannot be assessed on human subjects. There has been no relevant evidence to support whether the MADYMO human model can be used for forward fall investigations, which means that the course of forward fall event can be reconstructed using MADYMO human model. To our knowledge, this is the first study using a MADYMO human model for investigating forward falls.

## II. MATERIAL AND METHOD

In this study, we first conducted a series of laboratory experiments.

### A. Subjects

Four young Japanese individuals (male) of ages 20-25 years (mean ± standard deviation = 23.75 ± 0.5 years), participated in this study. Their average height and body mass were 174 ± 3 cm and 61 ± 5 kg, respectively. None of the participants reported any prior training, such as Jodo, martial arts techniques, wrestling, or gymnastics. Moreover, they did not have a history of major medical or neurological illnesses, such as epilepsy, hand trauma, balance disorders, and falls. The participants did not present bone disease or other conditions that would increase the likelihood of a fracture occurrence. All participants signed a written informed consent prior to the experiments. The experimental protocol was approved by the Institutional Review Board of Nagoya University, Japan. All experiments were conducted in accordance with the approved guidelines.

### B. Experimental Protocol

The objective of this experiment was to measure the applied forces to each hand during a forward fall. To secure the safety of the subjects and prevent injuries, the ground on which the experiment was conducted was covered with compliance flooring (i.e. a yoga mat).

The experiment was designed according to the protocol utilized in most prior forward fall studies [18]-[20]. This protocol was designed to be adapted in the *in vitro* study design [13]. In this study, each participant wore tight sportswear clothing and an upper body harness (Fig.1). The participants were instructed to place their knees on a soft pillow fixed to the ground and flex their knees, allowing the participant to lean forward against the harness. In this position, the thighs and the horizontal line parallel to the ground created an angle of 30° (Fig.1).

A rope connected to the harness lifted the participants to an appropriate height from the ground. The participants were instructed to maintain a position with their elbows fully extended (the outstretched hand position). In this position, the angle between the participants' arms and the vertical line was established as 15°. Two force plates (M3D Force plate, Tec Gihan Corp.) were fixed to the ground beneath the participants' hands. The surface of the force plates was considered to be a hard surface with infinite stiffness.

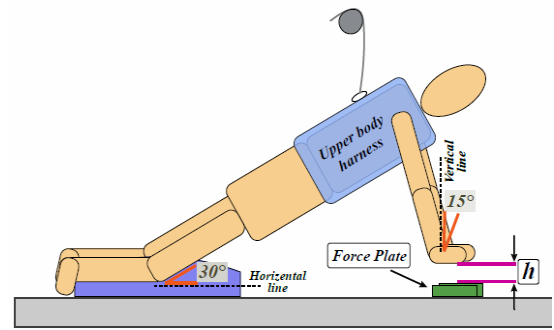

**Fig.1.** The experimental setup for the forward fall in the outstretched hand position for fall heights of 5 cm and 8 cm.

As previously mentioned, the participants' safety was the main concern during this experiment. In order to prevent a wrist fracture due to contact with the force plates, we designed the experiments considering the amount of impact force applied to each hand for two different fall heights ($h$ = 5 cm and 8 cm). The fall height was adjusted using the distance between the palms and the upper surface of force plates. Every participant completed six trials (two fall heights × three repetitions = six trials).

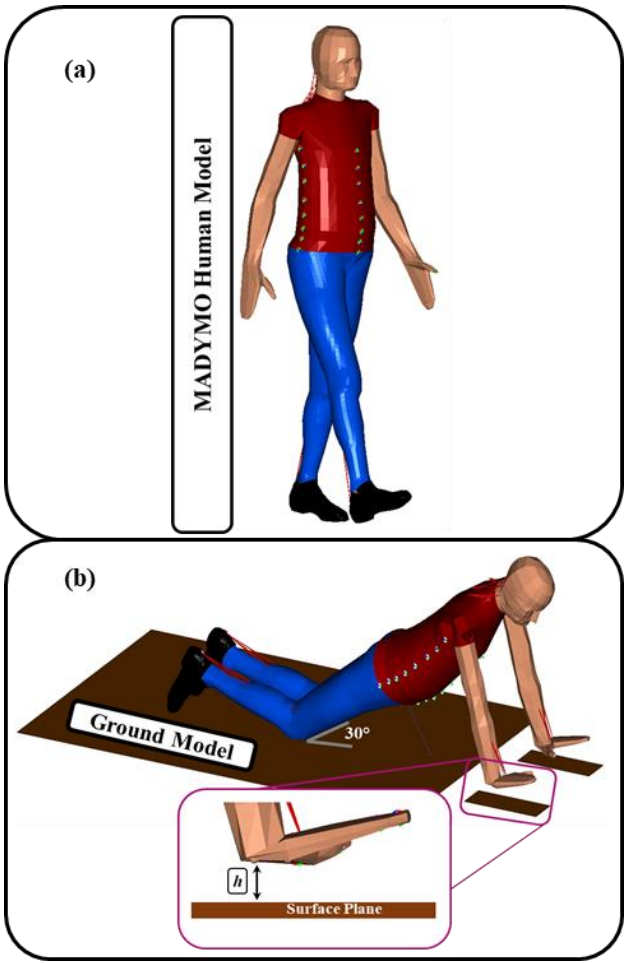

**Fig. 2.** The MADYMO Human model: (a) A mid-size male human model representing the 50th percentile in the standing position. (b) The MADYMO human model setup for simulating a forward fall. The fall height was adjusted based on the distance between the surface planes and the model's hands to h = 5 cm and 8 cm

*C. MADYMO Human Body Model Simulation Software*

In this study, we used an effective and extensive human model database, MADYMO, developed by TNO Automotive Safety Solutions (TASS) BV, Netherlands [25]. The MADYMO software consists of a solver, a database of human body models, and the pre- and post- processors. MADYMO has the capability to execute the numerical algorithm for predicting dynamic and kinematic behavior, such as representing the reaction motion of the human body. MADYMO is widely used in automotive occupant safety.

Moreover, this software has been utilized for other applications, such as the study of accident reconstruction, injury biomechanics, and sports medicine [27]-[28]. MADYMO also supports several databases of multibody human models. One of the released multibody human models is the pedestrian human body model (or human body model) [25]. The pedestrian model has a height 1.76 m and weight 75.3 kg and is released by default in a standing position. The model is based on the 50th percentile male model population, and its geometry was designed based on a database from the RAMSIS software package (RAMSIS 1997) [29].

In MADYMO, a multi-body system encompasses a system of rigid and flexible bodies that connect with kinematic joints (e.g., translational, sphere, or revolute joints). The model was designed based on 186 bodies, including 178 rigid and 8 flexible bodies. The chain body connections in the upper and lower extremities consist of several branches. Two separate branches of chain body connections were developed for connecting the pelvis to the vertebra and head, as well as connecting the feet to the left and right legs and shoes. The patella connects to the foot and toes with two discrete branches of chain bodies for the left and right foot. Similarly, two discrete branches were considered for constructing the left and right connections between the fingers and arms with the spine. Moreover, the soft tissue (flesh and skin) in the human model is characterized by force-penetrations to the 64 deformable ellipsoids chain bodies.

The model was verified by various loading tests, including a volunteer test for low severity loading and post modem human substitute tests for high severity loading [25].

*D. Reconstruction of the Forward Fall Scenario in MADYMO XMDgic Workspace*

In order to reconstruct the fall in a manner consistent with the experimental configuration (described in Section II.B), we modified the position of the MADYMO human model and created a ground surface in the MADYMO Xmagic workspace.

First, we modeled the ground using a plane surface with dimensions of 1.3 m × 1.0 m to be connected to the global system (reference space). The characteristics of the ground reaction force (force-penetration response) were defined such that the plane induced the reaction of a hard surface with infinite stiffness.

The released MADYMO human model defaults to a standing position, as shown in the Fig.2 (a); however, it is relatively simple to modify the dimensions and the posture. Therefore, the orientation of the kinematic joints for the ankles, knees, and hips were modified to generate a new human model position, as shown in the Fig.2 (b). In this new position, the angle between the knees and the X-Y plane was 30°. Moreover, we rotated the angle of the wrist joints to a value of 15° degrees between the wrists and X-Z plane.

In the MADYMO Xmagic workspace, we also designed two separate surface planes below the hands. The characteristics of force-penetration were considered, similar to the ground model. The use of discrete planes allowed us to adjust the distance between the hands and planes to represent the different fall heights. The fall height was adjusted to equal the distances used in the experiment ($h = 5$ cm and 8 cm).

The fall simulations were executed, while a load of 9.81 N (gravitational force) was applied to the human model. Each simulation time was established as 0.6 s, including the pre- and post-impact phases. The human body model reached the two ground planes in approximately 0.1 s and 0.13 s for the fall heights of 5 cm and 8 cm, respectively.

## III. RESULTS

The averages of the impact forces experienced by one hand are shown in Fig.3. The plots include the results of the human experiment (black-dash lines) and MADYMO simulation (blue-filled line) for the fall heights of 5 cm in Fig.3 (a) and 8 cm in Fig.3 (b). In addition, Fig.3 also includes the simulation results of a simple two-degree-of-freedom mathematical model (red-asterisk lines) represented in the previous study [18]. The model was designed to simulate a forward fall based on a spring-mass-damper system. In this model, the effective masses of the torso and upper extremities were assumed to be two separate masses. Moreover, the stiffness and damping of the shoulder-torso and wrist-palm were taken into consideration by including two separate springs and dampers.

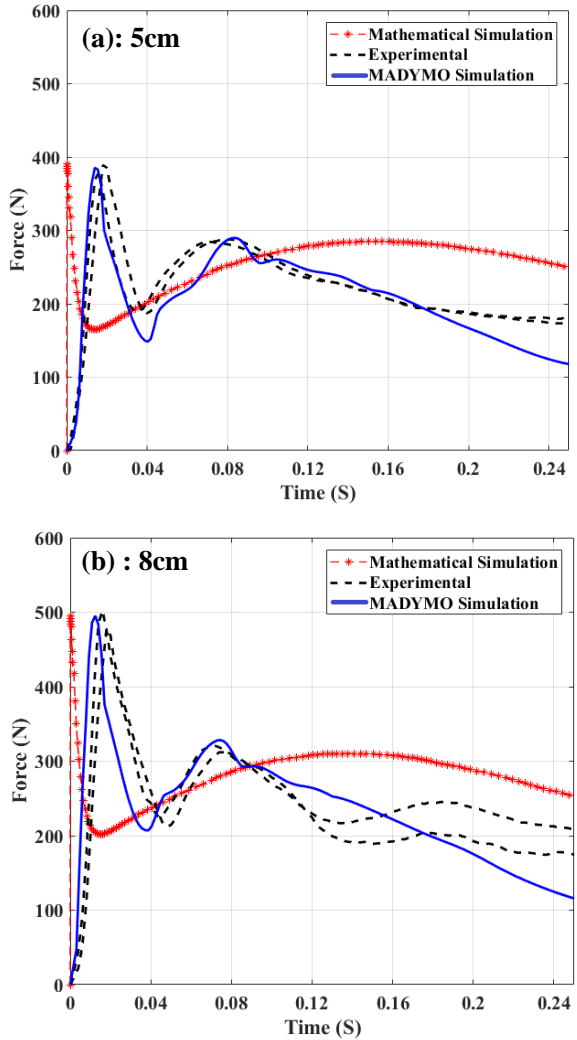

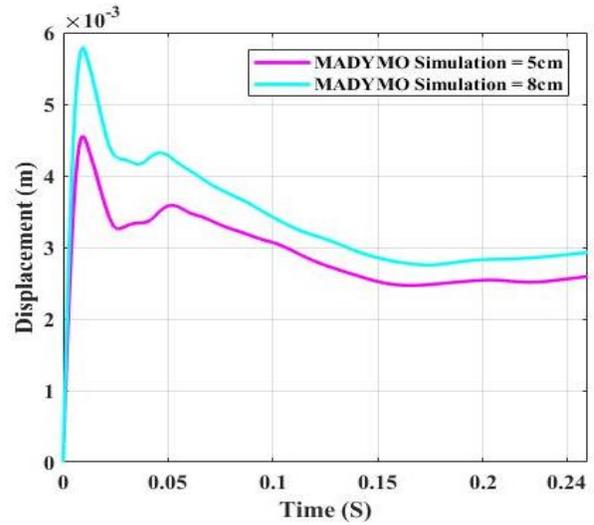

**Fig.4.** The MADYMO human model's hand displacement during forward fall from 5cm (magenta-filled line) and 8cm (cyan-filled line) fall heights.

**Fig.3.** The impact force profile during fall at (a) 5 cm and (b) 8 cm. The figures consists of experimental (black dash line), MADYMO simulation (blue filled-lien) and Mathematical model (red asterisk-line) results

Fig.3 demonstrates similar impact force characteristics between the human experiment, MADYMO simulation, and mathematical model. The impact forces include two peaks; the first one has a higher magnitude, which is followed by the lower second peak. The plots show that the MADYMO human model and mathematical model predict the magnitude of the two peaks with an accuracy corresponding to the experimental results. For the 5 cm fall, the first peak magnitudes were measured as 389 N, 385 N, and 391 N for the human experiment, MADYMO simulation, and mathematical model, respectively. The magnitudes of the second peaks demonstrated lower forces of approximately 286 N, 289 N, and 285 N for the human experiment, MADYMO simulation, and mathematical model, respectively. By increasing the fall height to 8 cm, we obtained the first impact magnitudes of approximately 490 N, 495 N, and 494 N for the human experiment, MADYMO simulation, and the mathematical model, respectively. Subsequently, we measured lower impact forces for the second peaks at 316 N, 328 N, and 310 N for the human experiment, MADYMO simulation, and mathematical model, respectively. The one-way analysis of variance (ANOVA), using the SPSS software package (IBM Corp. Released 2016. IBM SPSS Statistics for Windows, Version 23.0. Armonk, NY: IBM Corp), indicated that increasing the fall height significantly increases the magnitude of both peaks ($p < 0.0001$). Additionally, increasing the fall height had a greater effect on the first peak than the second peak.

The correlation coefficients for both fall heights were calculated to assess the relationship between the impact force for the human experiment and the MADYMO simulation. The results show a positive correlation; $r = 0.79$ for 5 cm and $r = 0.83$ for 8 cm between impact forces for the human experiment and the MADYMO simulation.

Fig.3 also shows that the first peak occurs approximately 0.025 s after the hand contacts the ground, while the second peak occurs at 0.075–0.08 s. The mathematical model tends to underpredict the time of the first peak and overpredict the time of the second peak [18].

Fig.4 shows the hand displacement in the MADYMO model during the fall for each of the two fall heights. During the 5 cm fall, the displacements for the first and second peaks were approximately 0.0046 m and 0.0036 m, respectively. By increasing the fall height to 8 cm, the displacements were approximately 0.0058 m and 0.0043 m for the first and second peaks, respectively. These results indicate that increasing fall height also has a significant impact on hand displacement ($p < 0.0001$).

## IV. Discussion

In biomechanical studies, the potential fall-related injuries on the upper extremity can be predicted using the mathematical model where *in vivo* and *in vitro* show ethical constraints [30]. In occupant safety studies, the use of human models is known as an alternative approach to investigate impact-related injuries during an accident [31]. Hence, in this study, we proposed to use one of the most advanced human body models to be developed by MADYMO software package [25]. We utilized MADYMO mid-size male model and investigated the applied force to the hand during a worst-case forward fall scenario (the outstretched hand position).

In the first step, we measured the magnitude of the impact force during the forward fall in the outstretched hand position through laboratory experiments. Two short distances were considered for the experimental fall heights ($h$ = 5 cm and 8 cm) due to avoiding the risk of participants' radius bone fracture. The results indicated that the impact force includes two distinct peaks, an initial high peak followed by a second lower peak. The results demonstrated that increasing the fall height, even for a very short distance, leads to an increase in the two peaks. Increasing the fall height has a significant effect on the first peak, leading to an increase of approximately 27%, while the second peak has an increase of only 10%. Consequently, our results provide additional evidence to support and confirm the previous findings regarding the impact force during a forward fall [18]-[22].

A similar experimental setup was applied to the MADYMO simulation. The position of the MADYMO human model was established to reflect that of the human experiment. The simulation was performed for the two fall heights when the model came into contact with the stiff surfaces. The simulation results for the impact forces indicated a similar trend with distinct peaks. Even though there is some discrepancy between the impact forces from the experiment and the MADYMO simulation, the MADYMO simulation results were acceptable and showed a positive correlation with the human experiment. There is a slight discrepancy between the magnitudes and times of the two peaks in the MADYMO simulation and the experiment. We speculate that this may be due to the algorithm used to solve the wrist and shoulder displacements and force productions.

In a previous study, a simple mathematical model based on a spring-mass-damper system with two degrees-of-freedom was proposed [18]. This model can accurately predict the magnitude of the impact forces (Fig.3), similar to the MADYMO human model; however, the time prediction of the first and second peaks was less accurate. The mathematical model tended to underpredict the time of the first peak (i.e. the first peak always occurs at the 0 s in Fig.3) and overpredict the time of the second impact force. This discrepancy in the time prediction is most likely related to the assumptions considered for the model. In other words, the model was designed with more attention given to predicting the peak of the impact forces rather than the impact time. In addition, the impact time prediction discrepancy suggests that the model is incapable of representing other factors, such as displacement and energy absorption. According to the model, the maximum hand displacements for the fall heights of 5 cm and 8 cm were approximately 0.153 m and 0.164 m, respectively. The maximum hand displacements in the MADYMO human model were approximately 0.0046 m and 0.0058 m during the 5 cm and 8 cm fall heights, respectively. Hence, the relatively large hand displacements from the mathematical model cannot accurately measure the amount of energy transferred to the hands during a fall from the standing position. However, the MADYMO human model seems to be capable of measuring the actual amount of energy absorption.

From the biomechanical perspective, the effective factors for determining the severity of the forward fall-related injuries are the peak of impact force and moment of impact. Hence, in this study, we demonstrate the MADYMO human model as a superior model compared to the prior mathematical model because it can accurately simulate the magnitude and time of the impact force, as well as the hand displacement. This study should prompt us to perform additional MADYMO simulations to investigate the injuries produced by a fall from the standing position.

## V. Conclusion

The main goal of this study was to evaluate the response of the MADYMO human model during a worst-case forward fall scenario (the outstretched hand position). Based on this objective, a set of human experiments was designed, and the impact forces applied to the hands for two short fall heights were measured. The short fall heights were considered due to avoiding the risk of participants' radius bone fracture. A similar setup was designed in the MADYMO workspace, and a simulation was performed. Our results demonstrated a strong correlation between the human experiment and the MADYMO human model, not only for measuring the magnitude of impact forces but also for providing accurate impact times.

Next, a comparison was carried out between the results of a mathematical model, which was designed based on a two degree-of-freedom spring-mass-damper system, and the MADYMO human model. Although both models accurately predicted the impact forces during the forward fall, the MADYMO simulation was more reliable, as the mathematical model indicated a discrepancy pertaining to the timing of the impact forces. This lack of reliability does not allow the model to measure hand displacement and energy absorption when the hands contact the ground. Because the hand displacement and energy absorption are critical factors for fall-related wrist fractures, this study recommends using the MADYMO human model for investigating a forward fall from a typical standing position.

## Acknowledgment

This work was supported by JSPS KAKENHI Grant Number 26750121, and METI "Strategic international standardization promotion project: International standardization of human tolerance against fall injuries".